\documentclass[a4paper,11pt]{article}
\usepackage{pos}
\usepackage{wrapfig}

\title{Direction Reconstruction using a CNN for GeV-Scale Neutrinos in IceCube}
 \ShortTitle{Neutrino Direction Reconstruction using a CNN}

\author{The IceCube Collaboration \\{\normalsize \normalfont(a complete list of authors can be found at the end of the proceedings)}}

\emailAdd{shiqi.yu@icecube.wisc.edu}

\abstract{The IceCube Neutrino Observatory observes neutrinos interacting deep within the South Pole ice. It consists of 5,160 digital optical modules, which are embedded within a cubic kilometer of ice, over depths of 1,450\,m to 2,450\,m. At the lower center of the array is the DeepCore subdetector. Its denser sensor configuration lowers the observable energy threshold to the GeV-scale, facilitating the study of atmospheric neutrino oscillations. The precise reconstruction of neutrino direction is critical in the measurements of oscillation parameters. This work presents a method to reconstruct the zenith angle of GeV-scale events in IceCube by using a convolutional neural network and compares the result to that of the current likelihood-based reconstruction algorithm.

\vspace{4mm}
{\bfseries Corresponding authors:}
Shiqi Yu$^{1*}$\\
{$^{1}$ \itshape Michigan State University}\\[4mm]
$^*$ Presenter

\FullConference{37$^{\rm{th}}$ International Cosmic Ray Conference (ICRC 2021)\\
		July 12th -- 23rd, 2021\\
		Online -- Berlin, Germany}

}

\begin{document}
\maketitle

\section{IceCube Neutrino Observatory}\label{sec:icecube}

\begin{figure}\centering
       \includegraphics[width=.7\textwidth]{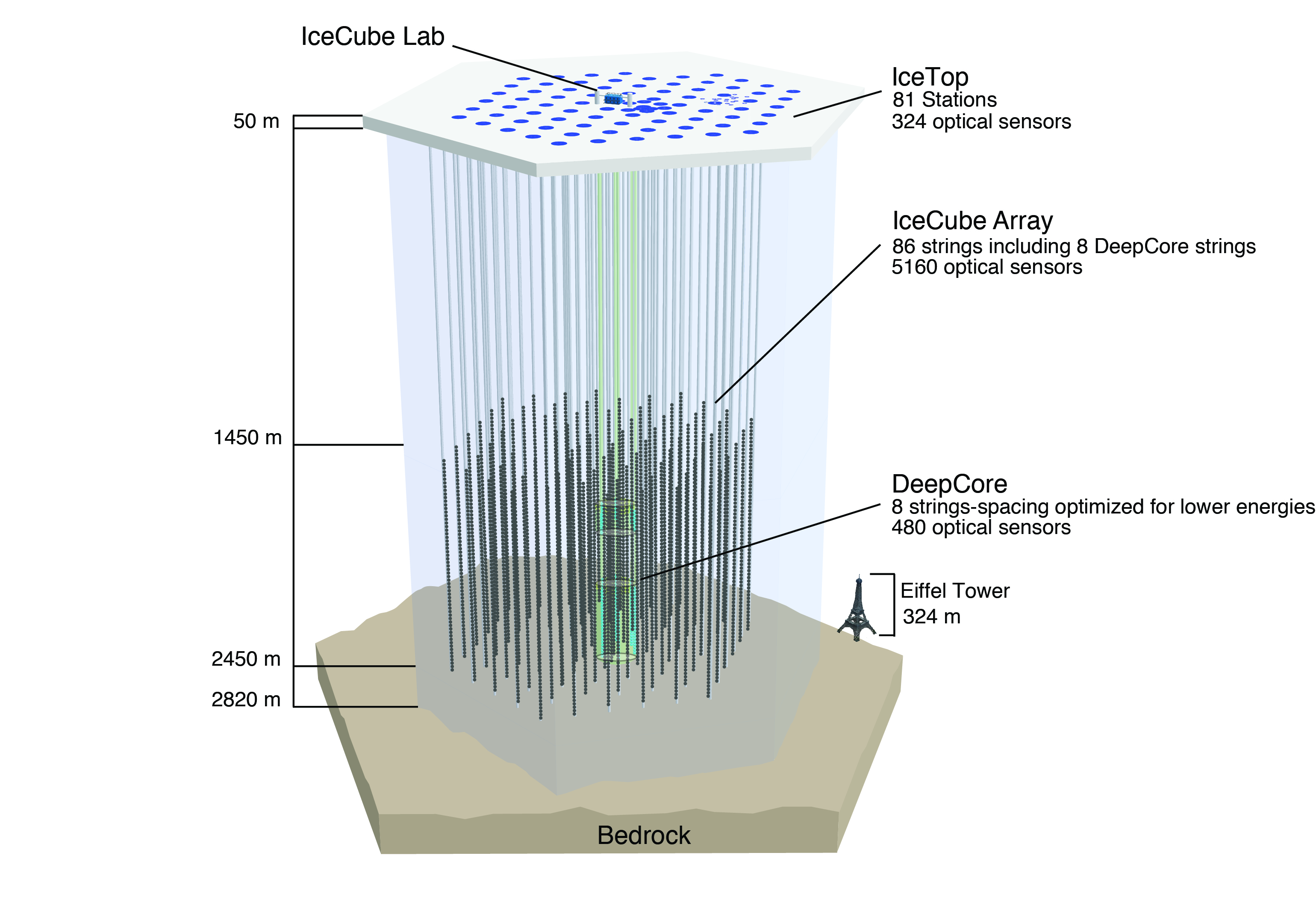}
       \caption{IceCube Neutrino Observatory at the South Pole}
       \label{fig:detector}
     \end{figure}

\noindent IceCube Neutrino Observatory is a Cherenkov detector located at the South Pole. As shown in Figure~\ref{fig:detector}, there are 5,160 digital optical modules (DOMs) deployed in the ice which make up of 78 IceCube strings and 8 DeepCore strings each containing 60 DOMs. The IceCube strings are arranged approximately 125\,m apart with the DOMs spacing as 17\,m. The DeepCore strings are located at the lower center of the IceCube string array with a denser configuration using DOMs with a (35\%) higher quantum efficiency.
The ten layers of DeepCore DOMs closest to the surface provide a veto on cosmic-ray muons which is an abundant background in the IceCube oscillation analyses. The DeepCore subdetector lowers the energy threshold from several TeV down to approximately 5\,GeV, allowing the study of neutrino oscillation in IceCube. 

\section{Neutrino Oscillation}\label{sec:icecube}
\noindent The DeepCore subdetector provides sensitivity to atmospheric mixing parameters,the mixing angle ($\theta_{23}$) and mass splitting ($\Delta m^2_{32}$). These can be measured by studying $\nu_\mu$ disappearance using atmospheric neutrinos that are created by cosmic rays interacting with the atmosphere.

Neutrinos are produced and detected as electron ($\nu_e$), muon ($\nu_\mu$), or tau ($\nu_\tau$) neutrinos, while they propagate in three mass eigenstates: $\nu_1$, $\nu_2$, and $\nu_3$. They can be produced in one flavor state but detected having a different flavor, which is called neutrino oscillations. $\nu_\mu$ disappearance is measured in the deficits of the $\nu_\mu\rightarrow\nu_\mu$ flux, the survival probability of which is described by
\begin{equation}P(\nu_\mu\rightarrow\nu_\mu)\approx 1-\sin^2(2\theta_{23})\sin^2(\frac{1.27\Delta m^2_{32}L}{E}),\end{equation}
where $L$ represents neutrino distance of travel; $E$ represents neutrino energy; and $\theta_{23}$ is the mixing angle and $\Delta m^2_{32} \equiv m^2_3 - m^2_2$ is the squared-mass difference between neutrino mass states $\nu_3$ and $\nu_2$. $L$ can not be directly known, but it can be inferred using incident neutrino zenith angle ($\theta_\text{zenith}$). 

When neutrinos interact within the detector, relativistic charged particles are produced, emitting Cherenkov photons which are detected by the DOMs and converted into series of electrical pulses. Precisely measuring neutrino $\theta_\text{zenith}$ is critical in measuring oscillation parameters. A convolutional network (CNN) is employed to reconstruct $\theta_\text{zenith}$ by using the series of electrical pulses of neutrino events.

\section{Method of Reconstruction}\label{sec:CNN}

\begin{figure}\centering
       \includegraphics[trim={1cm 5cm 1cm 7cm},clip, width=.6\textwidth]{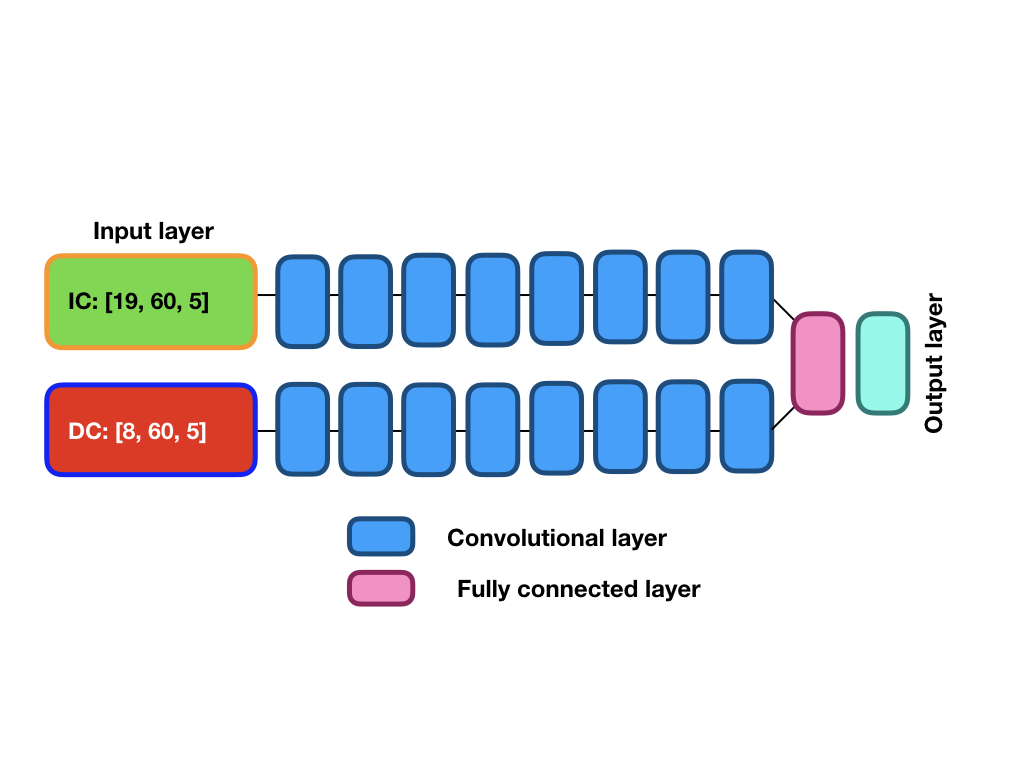}
       \caption{Structure of CNN with input shape of (number of strings, 60 DOMs, 5 variables), where DC represents 8 DeepCore strings and IC represents 19 nearby IceCube strings}
       \label{fig:cnn}
     \end{figure}
     
\begin{figure}\centering
       \includegraphics[width=.6\textwidth]{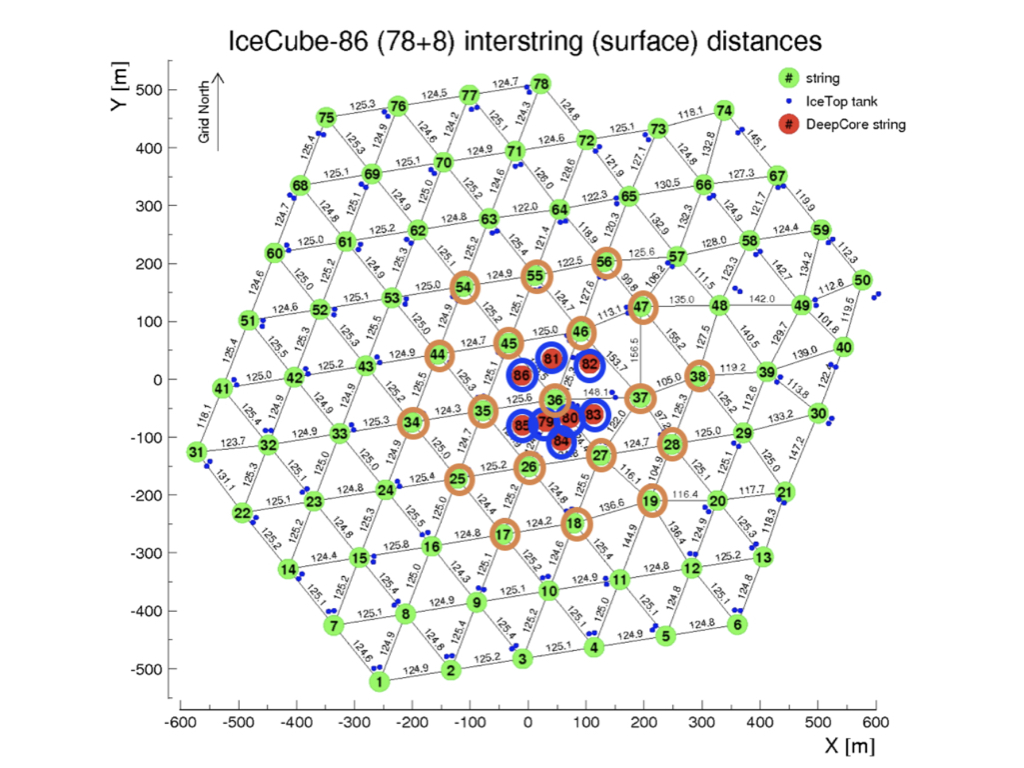}
       \caption{Top view of 8 DeepCore strings (red filled) and 19 IceCube strings (orange circled) used by CNN}
       \label{fig:icetop}
     \end{figure}
     
\noindent CNNs are broadly used in modern physics experiments for particle identification~\cite{paper:novacnn} and reconstruction~\cite{paper:dnnicecube}~\cite{paper:cnnenergy2021}. The CNN employed for $\theta_\text{zenith}$ reconstruction has the structure as shown in Figure~\ref{fig:cnn}. 

There are two sub-networks each of which consists of 8 convolutional layers. Each convolutional layer extracts some features from the input images and creates the output images which are used as the input to the following layer. The training samples are fed into the CNN via two input layers: one is for the 8 DeepCore strings; another is for the 19 surrounding IceCube strings, as shown in Figure~\ref{fig:icetop}. For each DOM on all the strings, 5 variables are calculated using the pulse series of the DOM: sum of charges, time of the first hit, time of the last hit, charge weighted mean of pulse time, and charge weighted standard deviation of pulse time. The output layer delivers the value of $\theta_\text{zenith}$ in the range of (0,$\pi$).

\begin{figure}\centering
       \includegraphics[width=.48\textwidth]{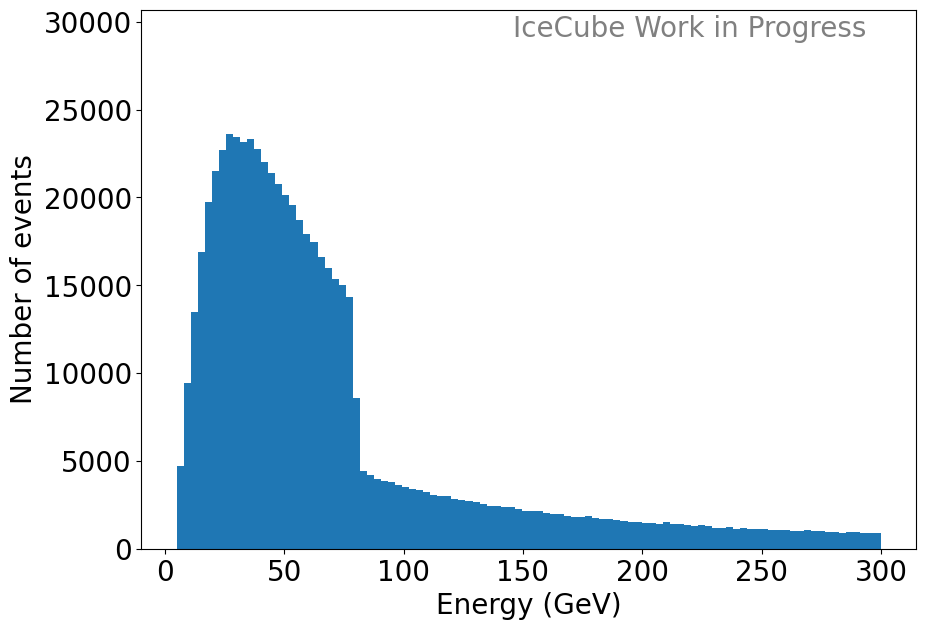}
       \includegraphics[width=.48\textwidth]{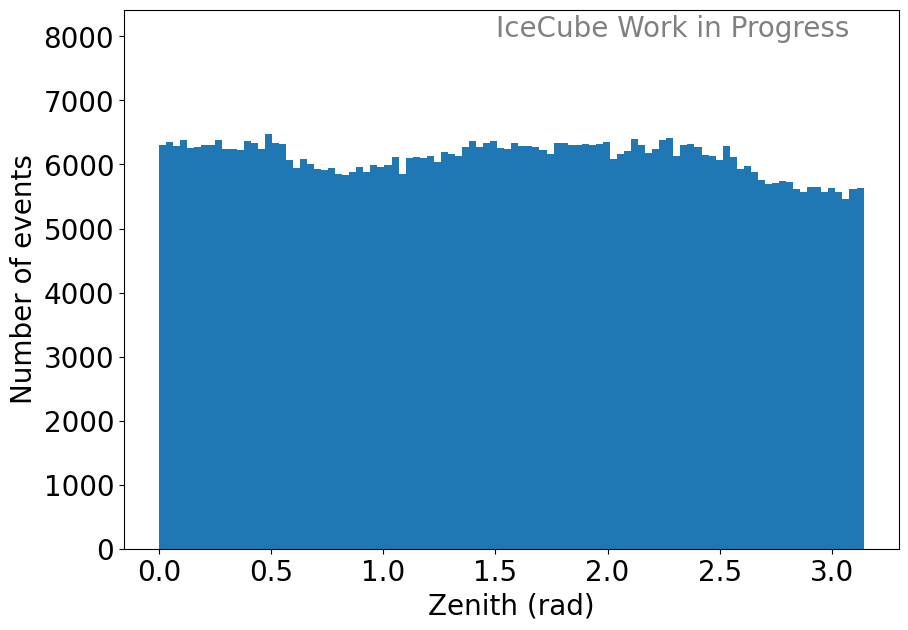}
       \caption{Unweighted true energy (left) and $\theta_\text{zenith}$ (right) distributions of training dataset}
       \label{fig:sample}
     \end{figure}

\begin{figure}\centering
       \includegraphics[width=.5\textwidth]{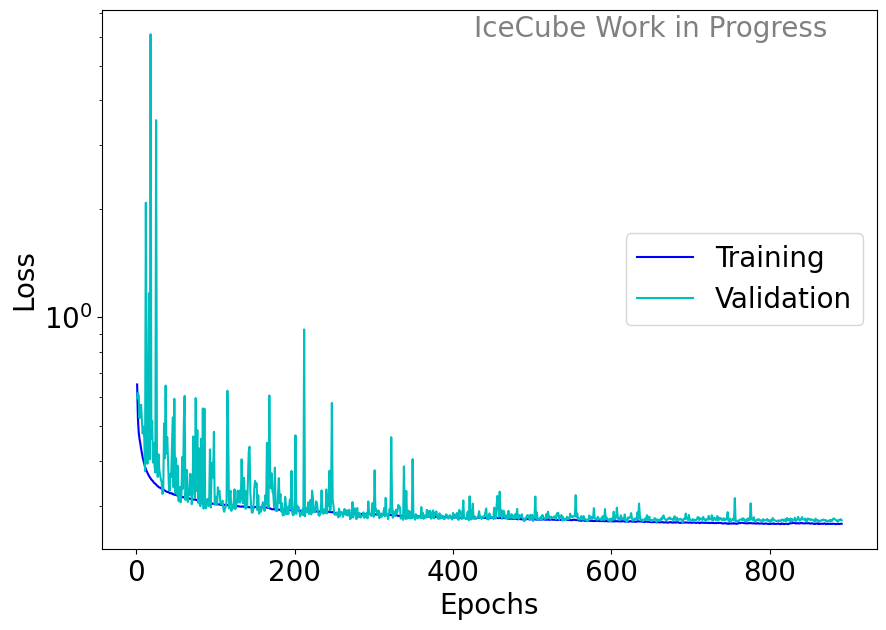}
       \caption{Training (blue) and validation (teal) loss curves}
       \label{fig:loss}
     \end{figure}
     
The training sample is simulated $\nu_\mu$ charged-current (CC) Monte-Carlo (MC) dataset with a flat $\theta_\text{zenith}$ distribution and energy between 5-300$\,$GeV, as shown in Figure~\ref{fig:sample}. A total of 5,024,876 simulated events were used to train the CNN, of which 80\% were used as a training set and 20\% for validation. The CNN was trained on the high performance computing at ICER, requiring approximately 6 days and over 800 epochs to converge. At the end of each epoch, the CNN updates all the parameters to minimize a loss function, defined as \begin{equation}
\text{loss}=\sum_{i}\hat{\theta}_\text{zenith}(i) - \theta_\text{zenith}(i),
\end{equation}
where $i$ represents event in validation dataset, $\hat{\theta}_\text{zenith}(i)$ represents the CNN predicted $\theta_\text{zenith}$ value of event $i$, and $\theta_\text{zenith}(i)$ represents the true $\theta_\text{zenith}$ value of event $i$. The training and validation loss curves are shown in Figure~\ref{fig:loss}.

\section{Results}\label{sec:results}

\noindent To show the performance of the CNN predicted $\theta_\text{zenith}$, a standard likelihood-based reconstruction method is used as comparison. The official $\nu_\mu$ and $\nu_e$ CC MC files are used to evaluate the results of these two reconstruction methods.
Selections based on the interacting point position (vertex) and energy of neutrino events that are reconstructed by the likelihood-based method are applied. These selections are inherited from the current oscNext analysis where they are optimized for neutrino oscillation signal efficiency. These likelihood-based selections are reconstructed: neutrino energy in range of [5, 300]GeV, $z$-coordinate of neutrino event vertex in range of [-500, -200]m, and $\rho_{36}<300\,$m, where $\rho_{36}$ represents radius of neutrino vertex relative to IC string 36. 
 
\subsection{$\nu_\mu$ CC sample}\label{subsec:numuresult}

\begin{figure}\centering
       \includegraphics[width=0.48\textwidth]{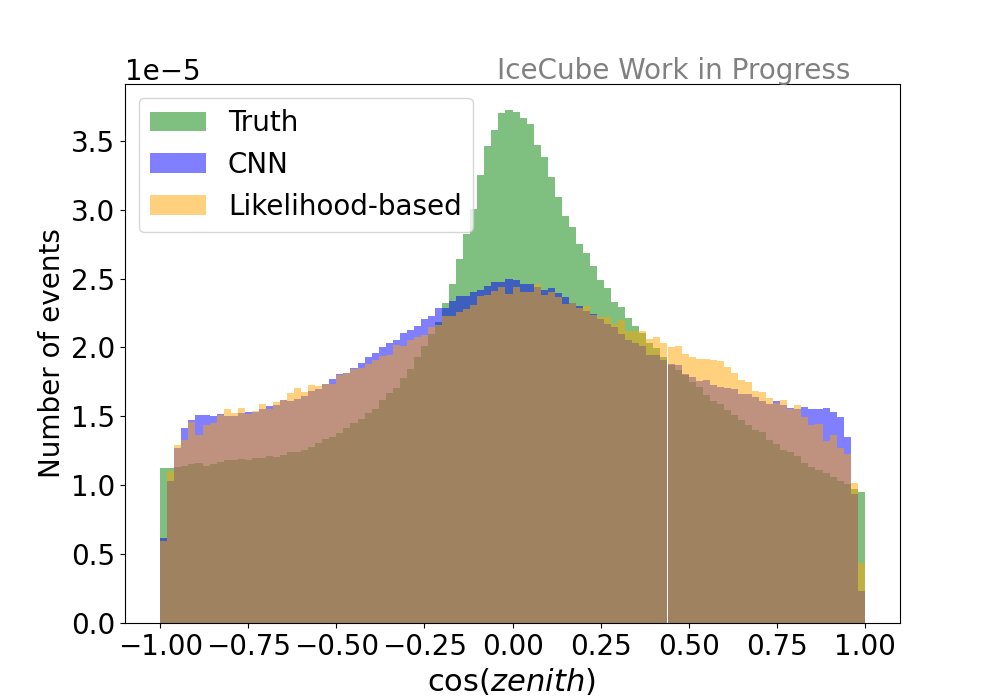}
       \includegraphics[width=.48\textwidth]{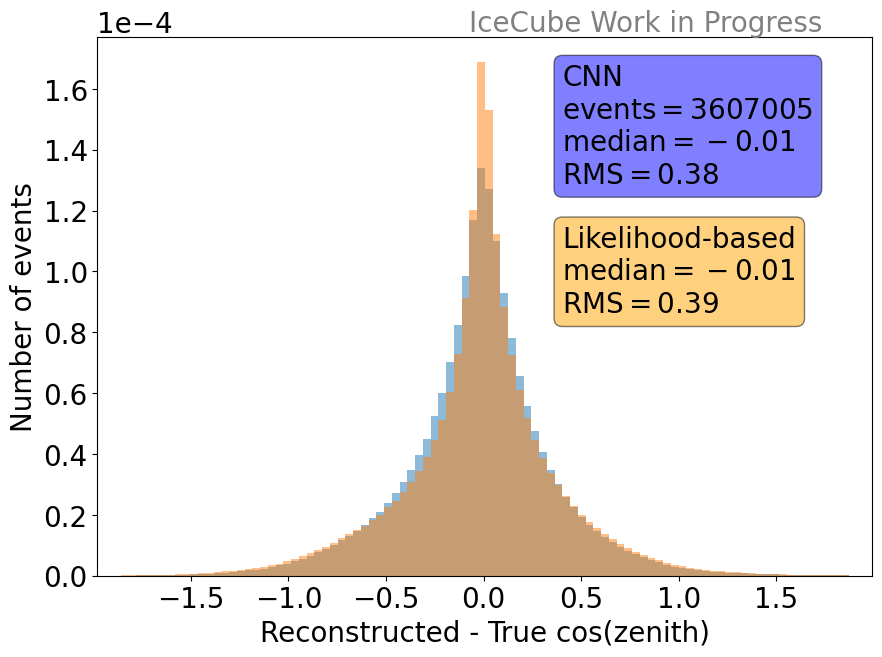}
       \caption{1D distributions of $\cos(\theta_\text{zenith})$ (left) and  $\cos(\theta_\text{zenith})$ reconstruction error (right) with blue (orange) representing CNN (likelihood-based) reconstructed $\cos(\theta_\text{zenith})$ and green representing true $\cos(\theta_\text{zenith})$ of true $\nu_\mu$ CC events.}
       \label{fig:numu1d}
     \end{figure}
\begin{figure}\centering
       \includegraphics[width=.48\textwidth]{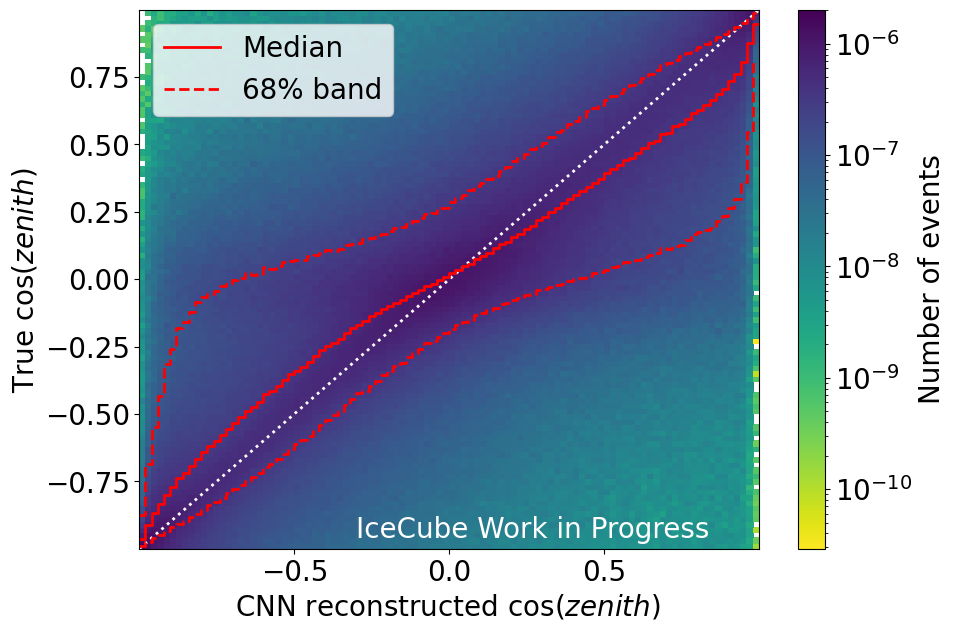}
       \includegraphics[width=.48\textwidth]{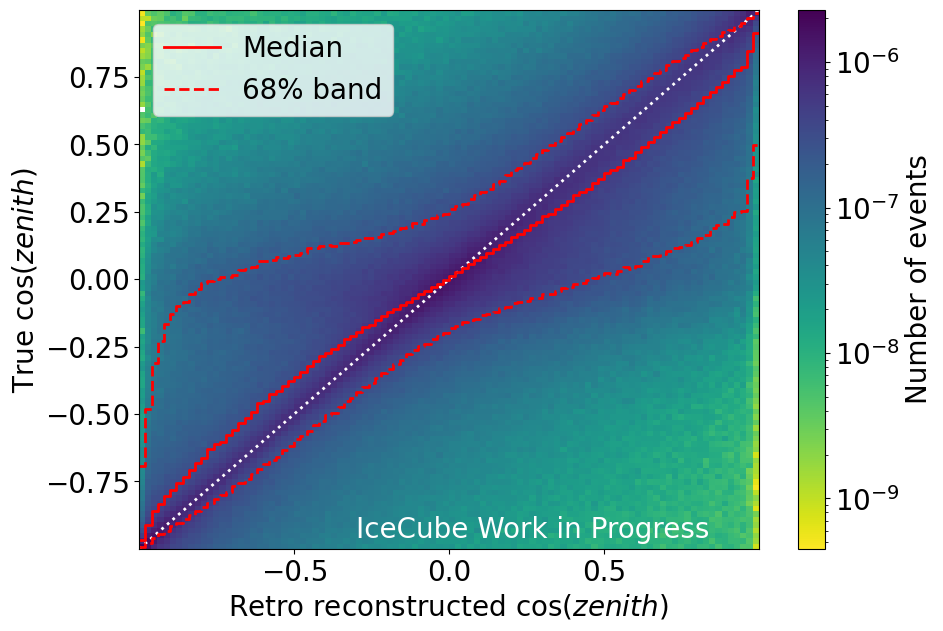}
       \caption{2D distributions of true vs.\ CNN (left) or likelihood-based (right) reconstructed $\cos(\theta_\text{zenith})$ with median (solid) and contours (dashed) of 68\% of events in vertical slices}
       \label{fig:numu2d}
     \end{figure}
     
\noindent The plots in Figure~\ref{fig:numu1d} show the 1D distributions of $\cos(\theta_\text{zenith})$. The CNN and likelihood-based methods have similar spectral shapes and both reconstructed $\cos(\theta_\text{zenith})$ values are smeared to the higher (lower) values at the lower (higher) boundary.

The 2D distributions of true versus reconstructed $\cos(\theta_\text{zenith})$ are shown in Figure~\ref{fig:numu2d}. Ideally, the median curve of the distribution should approach the diagonal white dot line, which represents the 1:1 ratio of true:reconstructed $\cos(\theta_\text{zenith})$ and the contours of 68\% of events should be narrowly parallel to the median curve. In the 2D distributions of the CNN and likelihood-based methods, both the medians and 68\%-contours are comparable.     

 \begin{figure}\centering
       \includegraphics[width=.48\textwidth]{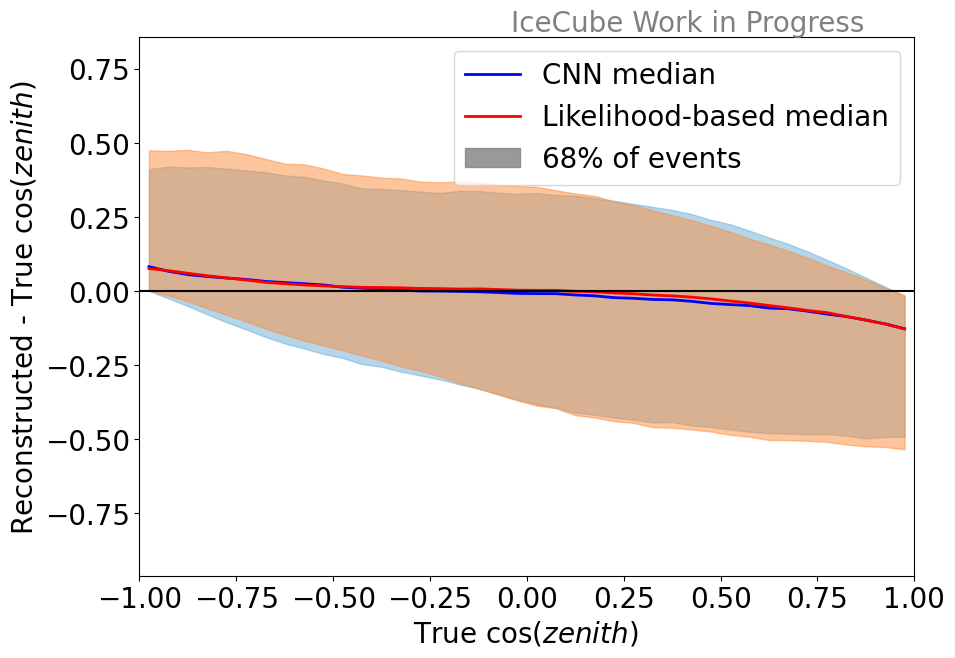}
       \includegraphics[width=.48\textwidth]{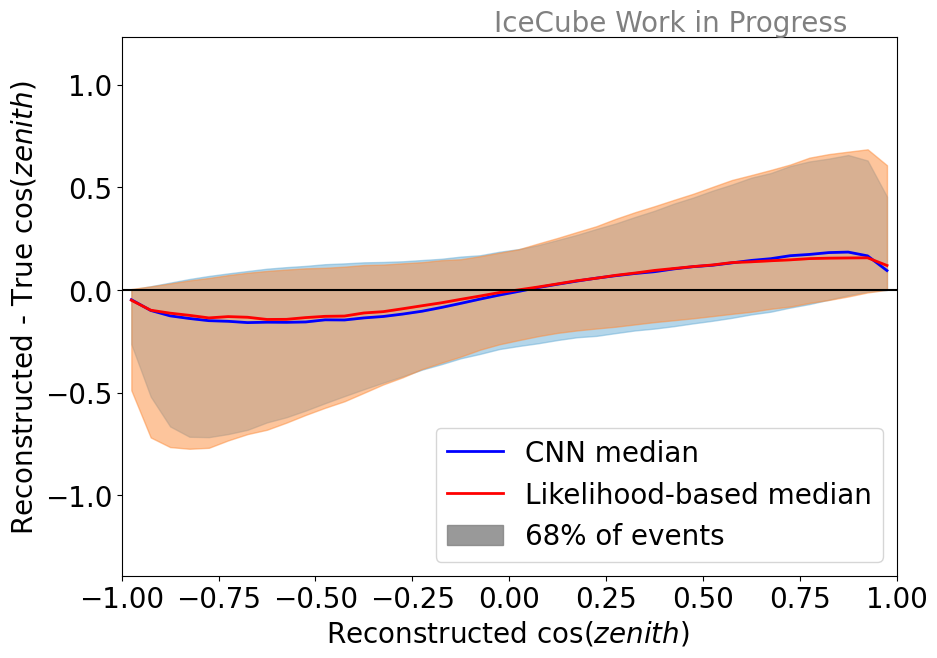}
       \caption{1D slices of reconstructed - true vs.\ true (left) or reconstructed (right) $\cos(\theta_\text{zenith})$ with blue (orange) representing CNN (likehood-based) result, solid curve representing median, and shaded area containing 68\% of events}
       \label{fig:numuslices}
     \end{figure}
     
As shown in Figure~\ref{fig:numu1d}, the overall RMS of CNN method is smaller than that of the likelihood-based method by 2.6\%. As shown in Figure~\ref{fig:numuslices}, plotting bias against true or reconstructed $\cos(\theta_\text{zenith})$ shows that the performances of CNN and likelihood-based methods are similarly well. 

\subsection{$\nu_e$ CC sample}\label{subsec:nueresult}

\begin{figure}\centering
       \includegraphics[width=0.48\textwidth]{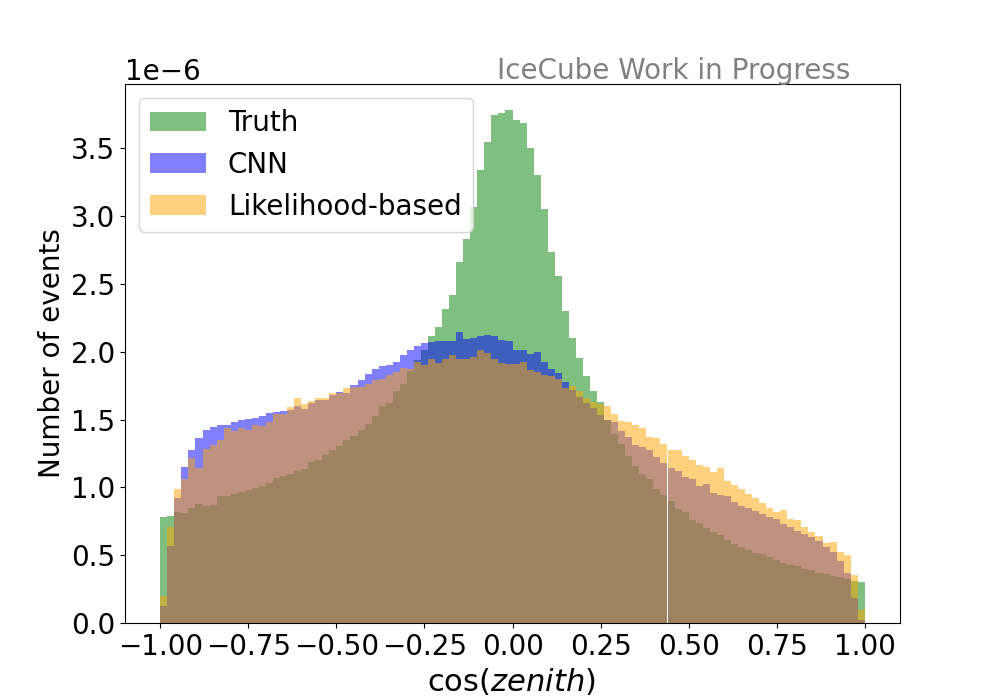}
       \includegraphics[width=.48\textwidth]{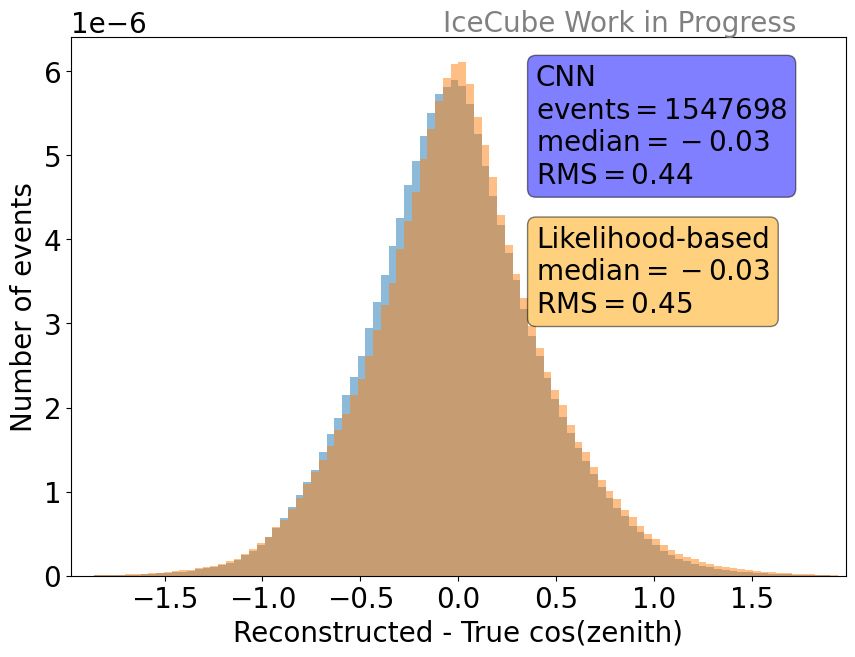}

       \caption{1D distributions of $\cos(\theta_\text{zenith})$ (left) reconstructed - true $\cos(\theta_\text{zenith})$ (right) with blue (orange) representing CNN (likelihood-based) reconstructed $\cos(\theta_\text{zenith})$ and green representing true $\cos(\theta_\text{zenith})$ of true $\nu_e$ CC events}
       \label{fig:nue1d}
     \end{figure}
\begin{figure}\centering
       \includegraphics[width=.48\textwidth]{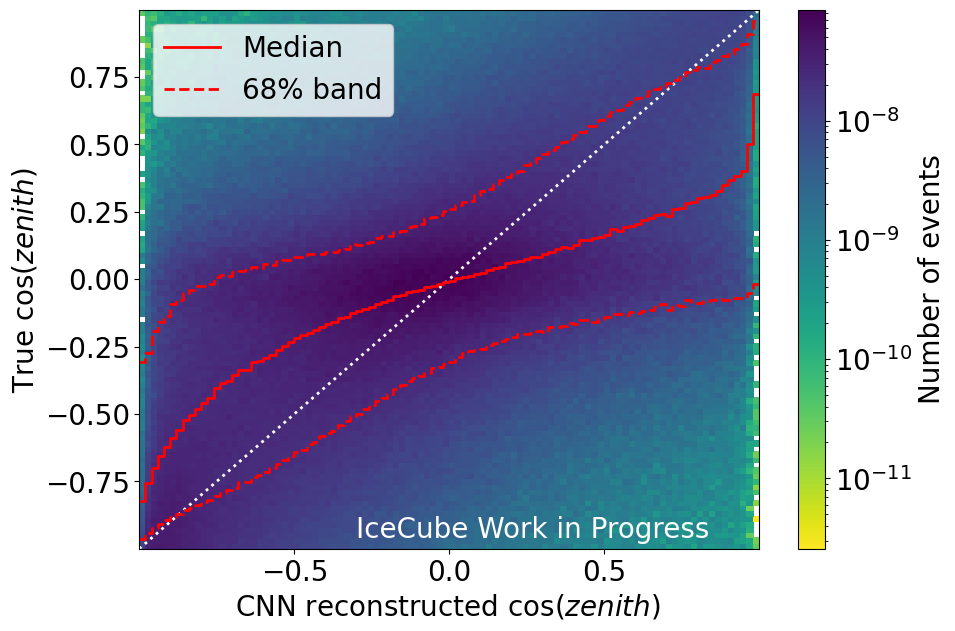}
       \includegraphics[width=.48\textwidth]{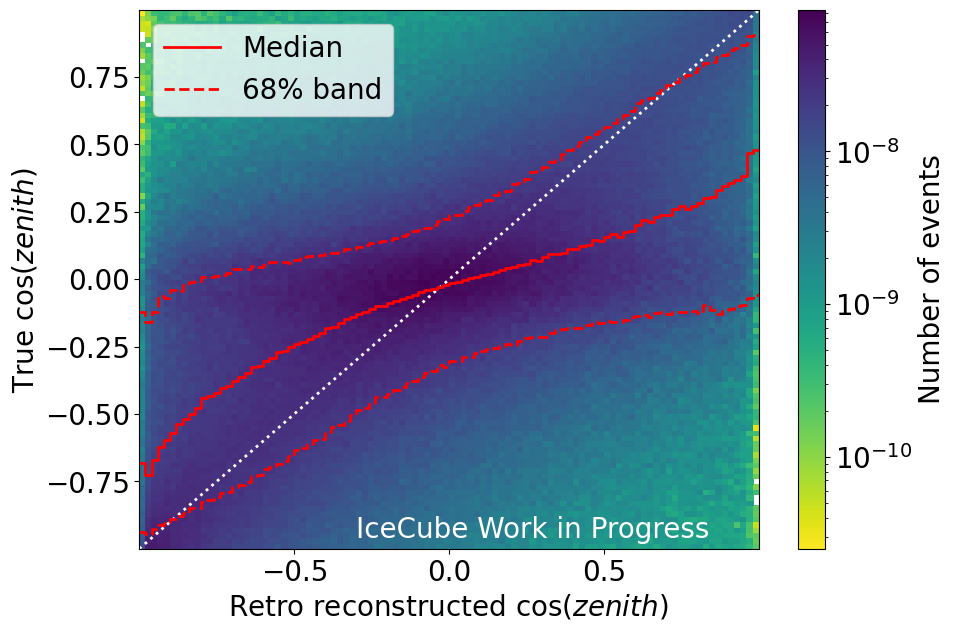}
       \caption{2D distributions of true vs.\ CNN (left) or likelihood-based (right) reconstructed $\cos(\theta_\text{zenith})$ with median (solid) and contours (dashed) of 68\% of events plotted on the top}
       \label{fig:nue2d}
     \end{figure}

\noindent In Figure~\ref{fig:nue1d}, the 1D distributions of $\cos(\theta_\text{zenith})$ of the CNN and likelihood-based methods have similar spectral shapes. The 2D distributions of true vs.\ reconstructed $\cos(\theta_\text{zenith})$ (see Figure~\ref{fig:nue2d}) look similar to each other while both having wider 68\%-contours than those of the $\nu_\mu$ CC events. This is as expected: most of the $\nu_\mu$ CC events are track-like in the IceCube detector and easier to reconstruct than the cascade-like $\nu_e$ CC events. This is also the reason that the bias distributions are wider and RMS values are larger in Figure~\ref{fig:nue1d} compared to those of the true $\nu_\mu$ CC events in Figure~\ref{fig:numu1d}.

 \begin{figure}\centering
       \includegraphics[width=.48\textwidth]{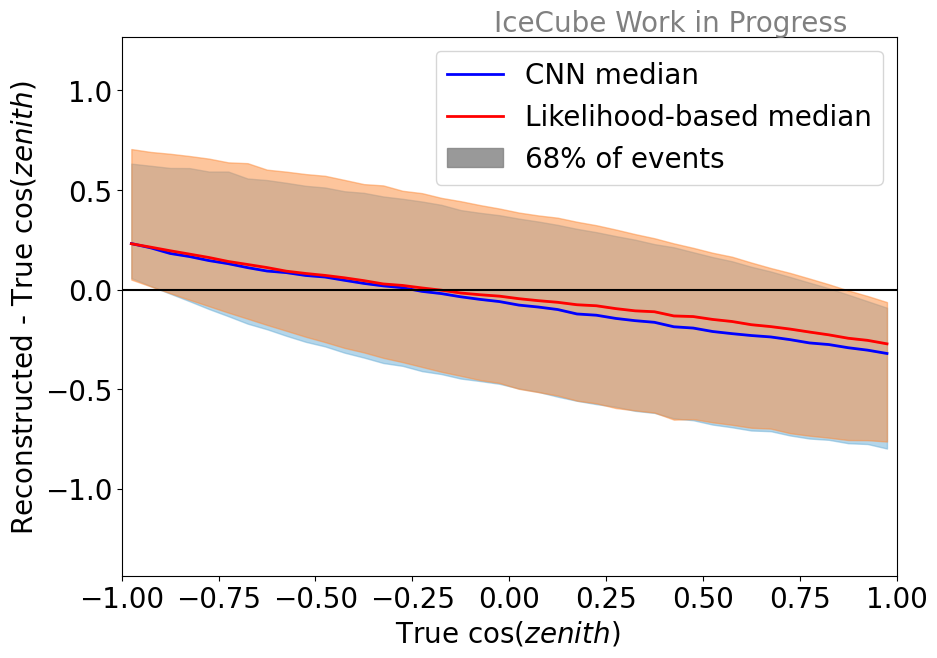}
       \includegraphics[width=.48\textwidth]{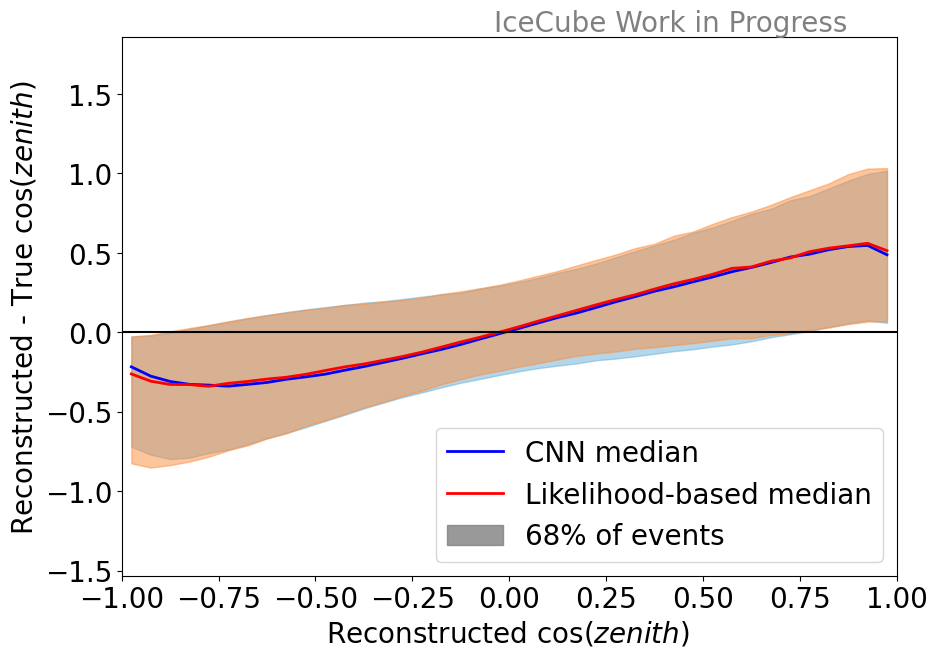}
       \caption{1D slices of reconstructed - true vs.\ true (left) or reconstructed (right) $\cos(\theta_\text{zenith})$ with blue (orange) representing CNN (likelihood-based) result, solid curve representing median, and shaded area containing 68\% of events}
       \label{fig:nueslices}
     \end{figure}
     
Similarly, as shown in the bias vs. true or reconstructed $\cos(\theta_\text{zenith})$ slices (see Figure~\ref{fig:nueslices}), the performance of the CNN method is comparable to that of the likelihood-based method, while both methods have worse performances than those of the $\nu_\mu$ CC events.

\subsection{Processing speed}\label{sec:speed}
\noindent As listed in Table~\ref{table:timing},  the likelihood-based method can only use CPU clusters while the CNN method can run on both CPU and GPU. The CNN benefits from parallel processing and is 10,000 times faster than the current method when run on a K80 GPU. Even if both methods are evaluated on the same CPU, running the CNN method is still over 400 times faster than running the likelihood-based method. Rapid processing is crucial for analyses using large data sets as is generally true of high energy physics experiments.


\begin{table}\centering
       \begin{tabular}{c|c|c}
           Second/Event & GPU & CPU \\
        \hline
        CNN & 0.0044  &  0.108 \\
        \hline
        Likelihood-based &  --   & 44.97  
       \end{tabular}
       \caption{Processing speed of CNN and likelihood-based methods}
       \label{table:timing}
\end{table}

\section{Conclusion}\label{sec:conclusion}
\noindent The CNN method trained on the simulated low-energy $\nu_\mu$ CC sample with a flat neutrino direction distribution provides a comparable performance to the current likelihood-based method, improving the overall RMS in the direction reconstruction by 2.6\% on the $\nu_\mu$ sample and 2.2\% on the $\nu_e$ events. The bias against either true or reconstructed $\cos(\theta_\text{zenith})$ slices are robust and comparable between the two methods. Using GPU resources, the CNN method is 10,000 times faster than the current method in processing events, easing the computational burden required for future oscillation analyses with DeepCore.

\section{Acknowledgement}\label{sec:acknowledgement}
\noindent This study is supported by National Science Foundation under grant number: NSF1913607. Any opinions, findings, or conclusions stated in this proceeding are only those of the author(s); and do not necessarily reflect the views of the National Science Foundation. Additional acknowledgements for the collaboration are included seperated. 


\bibliographystyle{ICRC}
\bibliography{references}

\clearpage
\section*{Full Author List: IceCube Collaboration}




\scriptsize
\noindent
R. Abbasi$^{17}$,
M. Ackermann$^{59}$,
J. Adams$^{18}$,
J. A. Aguilar$^{12}$,
M. Ahlers$^{22}$,
M. Ahrens$^{50}$,
C. Alispach$^{28}$,
A. A. Alves Jr.$^{31}$,
N. M. Amin$^{42}$,
R. An$^{14}$
% ...
\\

\noindent
% ...
$^{12}$ Universit{\'e} Libre de Bruxelles, Science Faculty CP230, B-1050 Brussels, Belgium \\
% ...
$^{14}$ Department of Physics and Laboratory for Particle Physics and Cosmology, Harvard University, Cambridge, MA 02138, USA \\
% ...
$^{17}$ Department of Physics, Loyola University Chicago, Chicago, IL 60660, USA \\
$^{18}$ Dept. of Physics and Astronomy, University of Canterbury, Private Bag 4800, Christchurch, New Zealand \\
% ...
$^{22}$ Niels Bohr Institute, University of Copenhagen, DK-2100 Copenhagen, Denmark \\
% ...
$^{28}$ D{\'e}partement de physique nucl{\'e}aire et corpusculaire, Universit{\'e} de Gen{\`e}ve, CH-1211 Gen{\`e}ve, Switzerland \\
% ...
$^{31}$ Karlsruhe Institute of Technology, Institute for Astroparticle Physics, D-76021 Karlsruhe, Germany  \\
% ...
$^{42}$ Bartol Research Institute and Dept. of Physics and Astronomy, University of Delaware, Newark, DE 19716, USA \\
% ...
$^{50}$ Oskar Klein Centre and Dept. of Physics, Stockholm University, SE-10691 Stockholm, Sweden \\
% ...
$^{59}$ DESY, D-15738 Zeuthen, Germany \\
% ...

\subsection*{Acknowledgements}

\noindent
USA {\textendash} U.S. National Science Foundation-Office of Polar Programs,
U.S. National Science Foundation-Physics Division,
U.S. National Science Foundation-EPSCoR,
Wisconsin Alumni Research Foundation,
Center for High Throughput Computing (CHTC) at the University of Wisconsin{\textendash}Madison,
Open Science Grid (OSG),
Extreme Science and Engineering Discovery Environment (XSEDE),
Frontera computing project at the Texas Advanced Computing Center,
U.S. Department of Energy-National Energy Research Scientific Computing Center,
Particle astrophysics research computing center at the University of Maryland,
Institute for Cyber-Enabled Research at Michigan State University,
and Astroparticle physics computational facility at Marquette University;
Belgium {\textendash} Funds for Scientific Research (FRS-FNRS and FWO),
FWO Odysseus and Big Science programmes,
and Belgian Federal Science Policy Office (Belspo);
Germany {\textendash} Bundesministerium f{\"u}r Bildung und Forschung (BMBF),
Deutsche Forschungsgemeinschaft (DFG),
Helmholtz Alliance for Astroparticle Physics (HAP),
Initiative and Networking Fund of the Helmholtz Association,
Deutsches Elektronen Synchrotron (DESY),
and High Performance Computing cluster of the RWTH Aachen;
Sweden {\textendash} Swedish Research Council,
Swedish Polar Research Secretariat,
Swedish National Infrastructure for Computing (SNIC),
and Knut and Alice Wallenberg Foundation;
Australia {\textendash} Australian Research Council;
Canada {\textendash} Natural Sciences and Engineering Research Council of Canada,
Calcul Qu{\'e}bec, Compute Ontario, Canada Foundation for Innovation, WestGrid, and Compute Canada;
Denmark {\textendash} Villum Fonden and Carlsberg Foundation;
New Zealand {\textendash} Marsden Fund;
Japan {\textendash} Japan Society for Promotion of Science (JSPS)
and Institute for Global Prominent Research (IGPR) of Chiba University;
Korea {\textendash} National Research Foundation of Korea (NRF);
Switzerland {\textendash} Swiss National Science Foundation (SNSF);
United Kingdom {\textendash} Department of Physics, University of Oxford.
\end{document}